\documentclass[twocolumn,pra]{revtex4}
\pdfoutput=1
\usepackage{amsfonts}
\usepackage[T1]{fontenc}
\usepackage{amsmath,amsbsy,amssymb,graphicx}
\usepackage{times}
\usepackage{color}

\newcommand{\td}{\mathrm{d}}
\newcommand{\be}{\begin{equation}}
\newcommand{\ee}{\end{equation}}
\newcommand{\bea}{\begin{align}}
\newcommand{\eea}{\end{align}}

\let\mathbf=\boldsymbol

\begin{document}

\title{{\Large Electrical detection of a skyrmion in a magnetic tunneling junction}}
\author{Keita Hamamoto$^1$ and Naoto Nagaosa$^{1,2}$}
\affiliation{$^1$Department of Applied Physics, University of Tokyo, 7-3-1 Hongo, Bunkyo-ku, Tokyo 113-8656, Japan}
\affiliation{$^2$RIKEN Center for Emergent Matter Science (CEMS), Wako, Saitama 351-0198, Japan}

\begin{abstract}
We theoretically investigated a method to detect a single skyrmion in a magnetic tunneling junction (MTJ) geometry. Using the tunneling Hamiltonian approach, we calculated the tunneling magnetoresistance (TMR) ratio of the skyrmion-ferromagnet bilayer system. 
%For wide range of parameters, the TMR ratio become larger than $30\%$, which is determined sorely by the spin profile of the skyrmion regardless of the electronic structures, if only the system is reasonably clean such that the spectral broadening is smaller than the exchange coupling between the local and the itinerant magnetic moment. 
We show the TMR ratio is determined sorely by the spin profile of the skyrmion and geometrical factor of the device, if only the system is reasonably clean such that the spectral broadening is smaller than the exchange coupling between the local and the itinerant magnetic moment. The TMR ratio in that case can amount to $30\%$ or higher when the diameter of the skyrmion is as large as the size of the device.
Since this criterion is easily achievable in real systems, MTJ geometry can be a good candidate of the electrical detection of a single skyrmion i.e., the reading process of the information in the future skyrmionics memory devices.
\end{abstract}

\maketitle

%{\it Introduction}: 
Magnetic skyrmion, a swirling texture of spins in magnets, is topologically protected particle-like object~\cite{nagaosa2013,fert2017}. Magnetic skyrmions appear in broad range of condensed matter systems, for instance, chiral magnets such as B20 compounds~\cite{yu2010,muhlbauer2009}, magnets with frustrated interactions~\cite{okubo2012,hayami2016,lin2016}, and interfaces of magnetic hetero-junctions~\cite{matsuno2016,dupe2016}. Due to its dynamical properties under relatively weak charge current~\cite{jonietz2010,iwasaki2013a} and the small size ($3\sim100 \ \mathrm{nm}$)~\cite{heinze2011}, applications to the low energy consumption memory devices are expected. Skyrmion racetrack memory~\cite{koshibae2015} is one of the most appealing examples, where the ferromagnetic domain walls carrying memory bits in original concept of racetrack memory~\cite{parkin2008} are replaced by skyrmions.

 For the implementation in devices, many efforts are made to create, delete, shift, and detect a single skyrmion.~\cite{koshibae2015,kang2016} Skyrmions are created/deleted by applying charge current~\cite{sampaio2013,romming2013}, local heating~\cite{finazzi2013,koshibae2014}, tailoring the device edge structure~\cite{iwasaki2013b}, and so on. Positional shift of skyrmions is realized by the current via the spin transfer torque quite effectively~\cite{jonietz2010,iwasaki2013a}. On the other hand, the detection of a single skyrmion is much more difficult. The Lorentz transmission microscopy~\cite{yu2010}, the spin-polarized~\cite{heinze2011} and the unpolarized scanning tunnel microscopy~\cite{crum2015,hanneken2015} can be used, however, the experimental equipment is quite expensive and the fast dynamics of individual skyrmion cannot be observed. To overcome these obstacles, a purely electrical method is highly desired. The detection of the number of skyrmions in a nano size Hall-bar device have been demonstrated using the topological Hall effect arising from the emergent magnetic field generated by the non-coplanar spin texture of skyrmions~\cite{kanazawa2015}. More recently, even the position of the individual skyrmion is theoretically proposed to be detected in similar nano device structure~\cite{hamamoto2016}.

In this paper we show another electrical detection method of a single skyrmion using the tunneling magnetoresistance (TMR) in a magnetic tunneling junction (MTJ) geometry. Since a skyrmion has many flipped spins near its center, where the electron tunneling is disturbed, one can easily expect that a skyrmion can affect the tunneling conductance. As the MTJ setup is intensively investigated in the long history of spintronics research, its implementation into the conventional devices, as well as into the future racetrack type devices will be feasible. 
By systematic calculations based on a simple model, we show that the TMR ratio can be larger than $30 \%$ for very wide range of parameters when the size of the skyrmion is comparable to that of the device. This result will pave a new way for the reading method of the information in future skyrmionics memory devices.

%{\it Model}: 
We consider a two-dimensional bilayer system consists of a skyrmion layer and a fully polarized ferromagnetic layer. The total Hamiltonian is the sum of the intra-layer double exchange model with the nearest neighbor hopping and the inter-layer tunneling Hamiltonian;
\begin{equation}
\hat{\mathcal{H}} = \hat{\mathcal{H}}_{sk} + \hat{\mathcal{H}}_F + \hat{\mathcal{H}}_T
\end{equation}
where
\begin{align}
\hat{\mathcal{H}}_{sk}&=-t\sum_{\langle ij\rangle \sigma }d^{\dagger}_{i\sigma}d_{j\sigma} -J\sum_{i\alpha\beta}d^{\dagger}_{i\alpha}\vec{\sigma}_{\alpha\beta}\cdot\vec{n}^{sk}_id_{i\beta} \\
\hat{\mathcal{H}}_F&=-t\sum_{\langle ij\rangle \sigma }c^{\dagger}_{i\sigma}c_{j\sigma} -J\sum_{i\alpha\beta}c^{\dagger}_{i\alpha}\vec{\sigma}_{\alpha\beta}\cdot\vec{n}^{F}_ic_{i\beta} \\
\hat{\mathcal{H}}_T&=-\sum_{ ij \sigma }T_{ij}c^{\dagger}_{i\sigma}d_{j\sigma} + h.c. .
\end{align}
$d^{\dagger}$ and $c^{\dagger}$ are the creation operators for each layer. The spin profile for the ferromagnetic layer is fixed as $\vec{n}^F_i =(0,0,1)$ while that for the skyrmion layer is assumed as $\vec{n}^{sk}_i = (\sin \theta_i \cos\phi_i,\sin\theta_i\sin\phi_i,\cos\theta_i)$ with $\theta_i=\pi (1-r_i/\lambda )$ for $r_i<\lambda$ and $\theta_i=0$ for $r_i>\lambda$ and $\phi_i=\varphi_i$. $r_i$ and $\varphi_i$ are the polar coordinate of two-dimensional plane and $\lambda$ is the radius of the skyrmion. As for the tunneling Hamiltonian, we only consider the vertical hopping $T_{ij}=T\delta_{ij}$ and we have neglected the spin flip tunneling. 

The tunneling current through the hetero-interface can be calculated by the standard perturbation theory with respect to the tunneling Hamiltonian~\cite{mahan}
\begin{align}
I_{sk} &= e\sum_{mn}\left|T_{mn}\right|^2 \int \frac{\mathrm{d}E}{2\pi} \left[ f(E)-f(E+eV)\right] \nonumber \\
 & \qquad \qquad \qquad \times A^{F}_m(E)A^{sk}_n(E+eV)
\end{align}
where $-e$ is the electron charge, $V$ is the voltage, $f$ is the Fermi distribution function, $T_{mn}$ is the tunneling Hamiltonian in the eigen basis of the both layers, and we have set $\hbar=1$. $A^{F/Sk}_m(E) = -2 \mathrm{Im} \left[ 1/(E-\varepsilon^{F/Sk}_m+i\Sigma) \right] $ is the spectral function for ferromagnetic/skyrmion layer with $\varepsilon^{F/Sk}_m$ being the $m$-th eigen energy of $\hat{\mathcal{H}}_{F/Sk}$ and $\Sigma$ is the spectral broadening which we is phenomenologically introduced to express both the elastic scattering by impurities and the inelastic scattering by electron-electron and electron-phonon scattering (at finite temperature). In this paper, we focus on the linear-response regime and the zero temperature limit;
\begin{align} \label{current}
I_{sk} &= \frac{e^2V}{2\pi}\sum_{mn}\left|T_{mn}\right|^2 \frac{2\Sigma}{(\varepsilon^F_m-\mu)^2+\Sigma^2}\frac{2\Sigma}{(\varepsilon^{sk}_n-\mu)^2+\Sigma^2}
\end{align}
where $\mu$ is the chemical potential.

The TMR ratio is defined as
\begin{equation}
\mathrm{TMR}\equiv \frac{R_{sk}-R_P}{R_P} = \frac{I_P}{I_{sk}}-1
\end{equation}
with $R_X=V/I_X$ being resistances for parallel ferromagnet-ferromagnet (X=P) or skyrmion-ferromagnet (X=sk) configuration. In this formalism, the tunneling amplitude $|T|$ does not affect the TMR ratio.

%{\it Conventional estimations of TMR ratio}:
TMR ratio in the anti-parallel configuration is usually estimated using spin polarization $P$ or the spin-resolved density of states (DoS) $D_\sigma$ as ~\cite{maekawa1982}
\begin{equation}
\mathrm{TMR}\sim\frac{2P^2}{1-P^2}
\end{equation}
with $P=\left(D_\uparrow -D_\downarrow\right)/\left(D_\uparrow +D_\downarrow\right)$. This estimation is equivalent to the assumption $I\sim D_\uparrow D_\uparrow +D_\downarrow D_\downarrow$ which is obtained simply putting $T_{mn}=T$ in eq.~(\ref{current}).
One can easily generalize this estimation to the cases in the MTJ with inhomogeneous magnetic structure such as the skyrmion system by using the local density of states (LDoS) as $ I\sim\sum_i\left( D_{i\uparrow} D_{i\uparrow} +D_{i\downarrow} D_{i\downarrow} \right)$. Another estimation, which is the simplest one neglecting the electronic structures, is the sum of the inner product of local magnetic moments; $I\sim \sum_{i}\frac{1+\vec{n}_i^{sk}\cdot\vec{n}_i^{F}}{2}=\sum_{i}\cos^2\frac{\theta_i}{2}$. This estimation is just counting the effective number of passable sites where vertical pairs of local spins are parallel. In the following, we show that the faithfully calculated TMR ratio using eq.~(\ref{current}) reduces to the value of the final simplest estimation and clarify its conditions. In the calculations below, we set the size of the two-dimensional layer as $L^2=100^2$ in the unit of the lattice constant and the radius of the skyrmion is $\lambda=50$, the maximum size in the square-shaped device, until otherwise specified.

\begin{figure}[t]
\centerline{\includegraphics[width=0.5\textwidth]{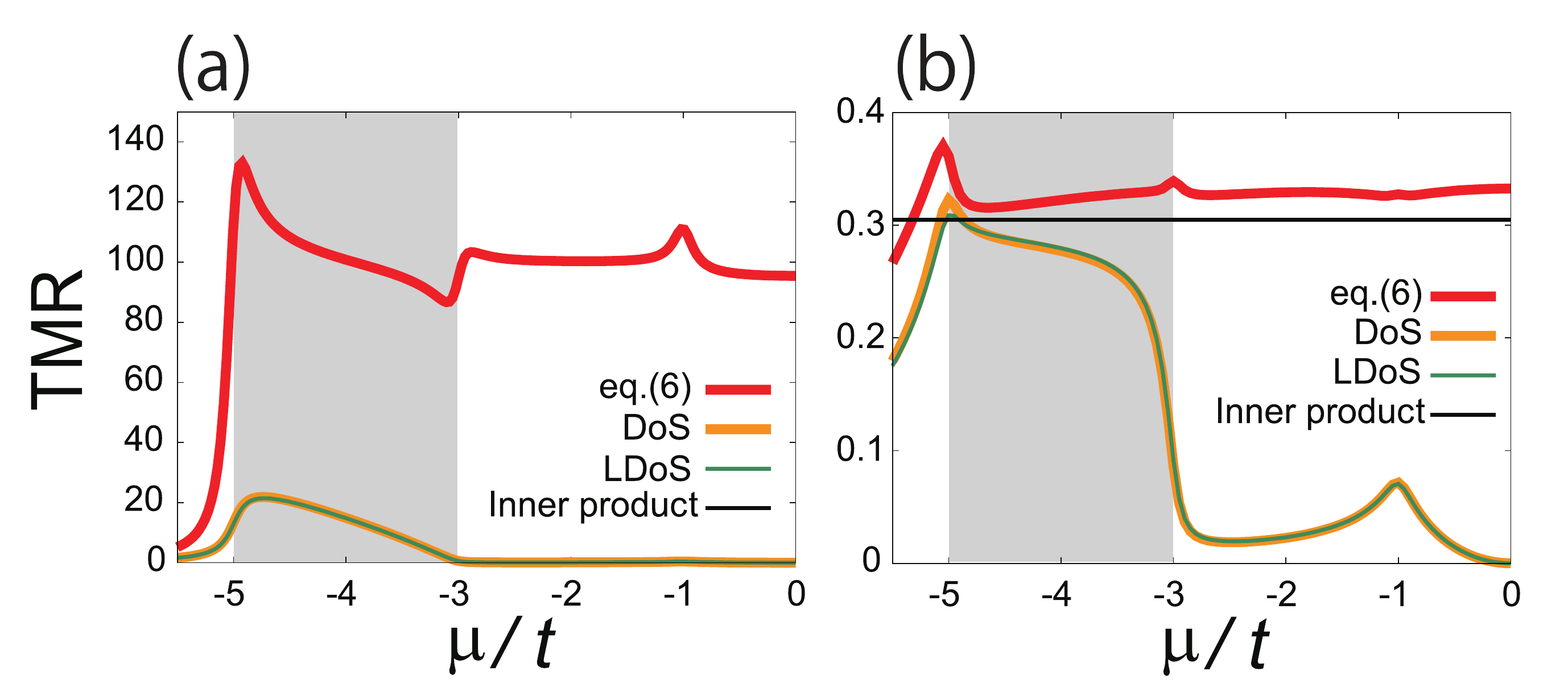}}
\caption{ TMR ratios for (a) the anti-parallel ferromagnet-ferromagnet, and (b) the skyrmion-ferromagnet bilayer system. Parameters are $J=t,$ and $ \Sigma=0.1t $. Red line is the TMR ratio calculated using eq.~(\ref{current}). Estimation of the TMR ratio from the density of states (yellow line), and the local density of states (thin green line) increases in the half-metallic (shaded) region. The black line in (a) cannot be seen since it is infinity. }
\label{FigMudep}
\end{figure}

%{\it Results}: 
Figures~\ref{FigMudep}(a) and (b) show the chemical potential dependence of the TMR ratios in the anti-parallel ferromagnet-ferromagnet and the skyrmion-ferromagnet bilayer system, respectively. Parameters are set as $J=t,$ and $\Sigma=0.1t$. The red line is calculated from eq.~(\ref{current}), and yellow, green, black lines are estimations from DoS, LDoS and the inner product of the local magnetic moments, respectively. The black line in Fig.~\ref{FigMudep}(a) cannot be seen since it is infinity. On the other hand, in the case of skyrmion-ferromagnet bilayer case in Fig.~\ref{FigMudep}(b), the inner-product describes well the obtained TMR ratio (red line). TMR ratio estimated from DoS (yellow line) coincide well with that from LDoS (green line), which indicates spatial inhomogeneity of the electronic structure does not affect the TMR ratio. This result is in sharp contrast with previous studies based on LDoS description of the tunneling conductance in skyrmionic systems~\cite{crum2015,hanneken2015}. We will discuss on this difference later in detail.
  For $-4t-J < \mu < -4t+J$, (shaded region in Fig.~\ref{FigMudep}(b)) the system is half-metallic i.e. DoS of minority spin vanishes. In this region, the estimation from DoS approaches the real value (red line), but suddenly decrease in the non half-metallic regime. Nevertheless, the real TMR ratio is not so sensitive on weather the system is half-metallic or not. This energy independence indicate the finite temperature effects shall be quite small. Hereafter we focus on the energy window $-4t+J+0.1t < \mu < -4t+J+0.5t$, which is non half-metallic region for small $J$, and we will show averaged values of TMR ratio within this window. 
The standard deviation is not shown since it is so small that we cannot see, excepting the small $\Sigma$ region where the discretized nature of energy levels appears due to the finite size effect as commented below. All the following results are qualitatively the same for the half-metallic regime.

\begin{figure}[!t]
\centerline{\includegraphics[width=0.5\textwidth]{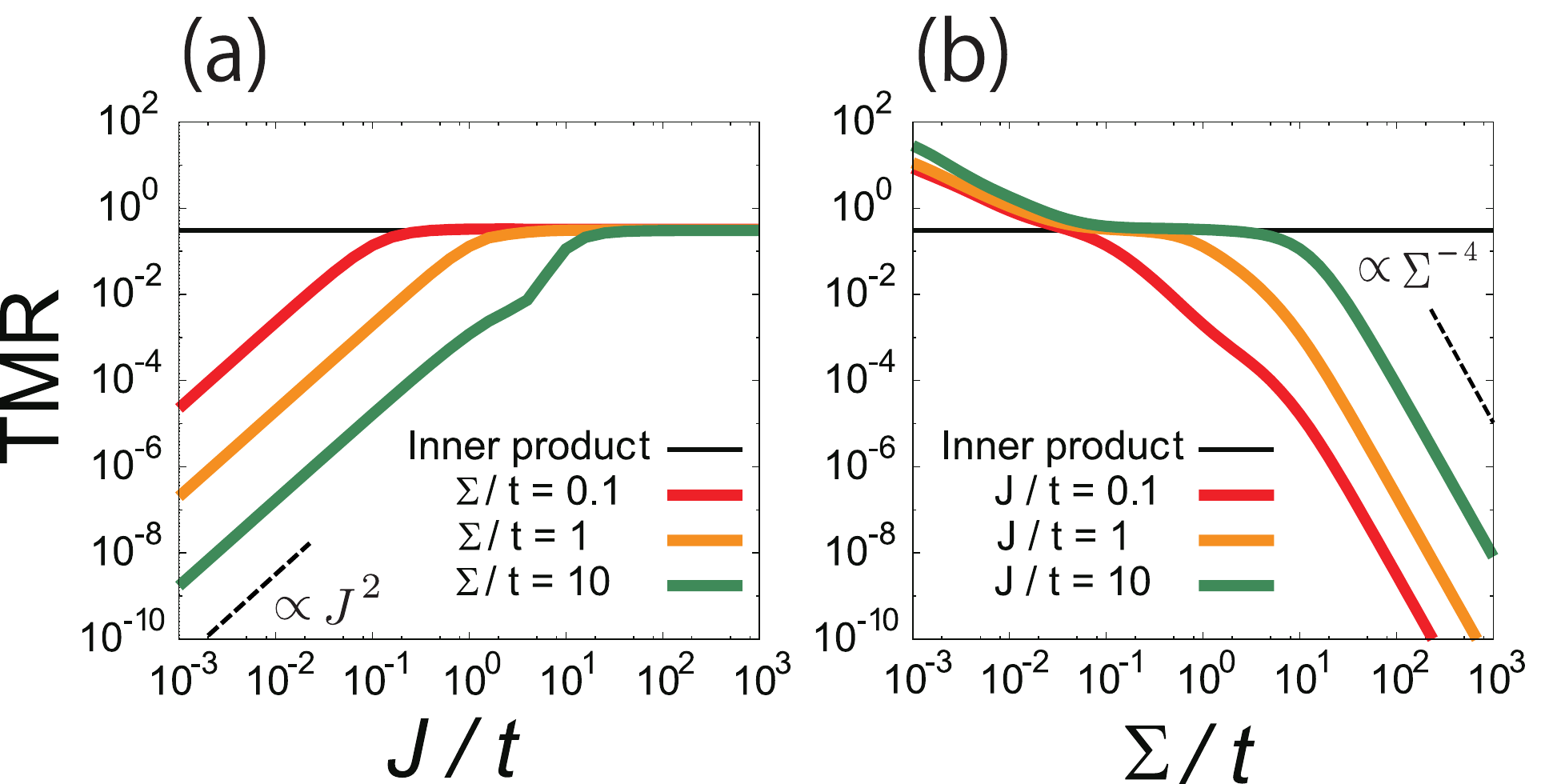}}
\caption{ Log-log plot of (a) $J$ dependence of the TMR ratio of skyrmion-ferromagnet bilayer system for various values of $\Sigma$ and (b) its $\Sigma$ dependence for various values of $J$. The black line is the estimated TMR ratio from the inner product of the local magnetic moments. All the results converge to this black line when $t \ll \Sigma\ll J$ is satisfied.}
\label{FigSigmaJdep}
\end{figure}

 Figure~\ref{FigSigmaJdep}(a) shows the $J$ dependence of the TMR ratio in the skyrmion-ferromagnet bilayer system. We can clearly see the relation $\mathrm{TMR} \propto J^2$ for small $J$ region. This exponent is understood in terms of the perturbative calculation with respect to $J$ as shown in the Supplementary Materials (S1). The TMR ratio converges for larger $J$ to the value estimated from the inner product of the local magnetic moments (black line). 

The $\Sigma$ dependence is shown in Fig.~\ref{FigSigmaJdep}(b). For the clean system, due to the discreteness of the energy spectrum, the TMR ratio strongly depends on the chemical potential $\mu$. In this case, we could not specify the exponent of the divergence of the TMR ratio in the limit of $\Sigma\rightarrow 0$. However we can speculate the exponent of the divergence as $\mathrm{TMR} \propto \Sigma^{-2} \sim \tau ^2$ with $\tau$ being the transport lifetime of the electron. Detailed discussion is given in the Supplementary Materials (S2).
% the TMR ratio shows usual diffusive transport nature; $\mathrm{TMR}\propto\Sigma^{-1}\sim\tau$ with $\tau$ being the transport lifetime of the electron. 
For the disordered case $\Sigma \gg J$, the TMR ratio is proportional to $\Sigma^{-4}\sim\tau^4$. In this region, even the half-metallic nature is completely smeared out. We can analytically prove this $\Sigma^{-4}$ dependence from eq.~(\ref{current}). Details are given in the Supplementary Materials (S3).
For the intermediate region, the TMR ratio becomes the same value as that estimated from the inner product of the local magnetic moments ($\sim 30 \% $ for $\lambda=50$) and independent of $\Sigma$. This plateau region expands for larger $J$. In this regime, namely $t\ll  \Sigma \ll J$, the spins of itinerant electrons are forced to align to the local magnetic moments. The transfer integrals therefore have to include the factor of the overlap of wave functions in the spin space; $t_{ij}=t\rightarrow t\langle i|j\rangle  = t e^{i a_{ij}}\cos\frac{\theta_{ij}}{2} $ where $a_{ij}$ is the emergent gauge field which accounts for the emergent magnetic field arising from the non-coplanar spin texture, and $\theta_{ij}$ is the angle between local magnetic moments $\vec{n}_i$ and $\vec{n}_j$. The inter-layer tunneling amplitude is also modified as $T_{ij} = T\delta_{ij} \rightarrow T \delta_{ij}e^{ia_i}\cos \frac{\theta_i}{2}$. In this situation, the system is spin-less and parameter $J$ does not enter into the Hamiltonian nor the eigenenergies explicitly. When the spectral broadening $\Sigma$ is much larger than $t$ in addition, the tunneling current in eq.~(\ref{current}) is reduced to $I=4e^2V \Sigma^{-2} \sum_{mn} |T_{mn}|^2\propto \mathrm{Tr}[TT^\dagger]=  \sum_{i}\cos^2\frac{\theta_i}{2}$. Since the prefactor cancels out, the TMR ratio becomes the same value as evaluated only from the local magnetic moments and independent of $\Sigma$. Physically, the electronic structure for each spin is smeared out due to the large $\Sigma$, but different spin states do not mix since $\Sigma \ll J$. In such situation, the detailed electronic structures cannot play any roles but the spin polarization survives, therefore, the simplest estimation taking into account only the spin information, neglecting the electronic structure, gives us a good agreement. 

In summary, the TMR ratio is larger than $\sim 30 \%$ if only $\Sigma \lesssim J$ and $\lambda\simeq L/2$ is satisfied. This criterion is simple and experimentally feasible, and we can conclude that our setup can be a good candidate for the skyrmion detection in MTJ devices.

\begin{figure}[!t]
\centerline{\includegraphics[width=0.4\textwidth]{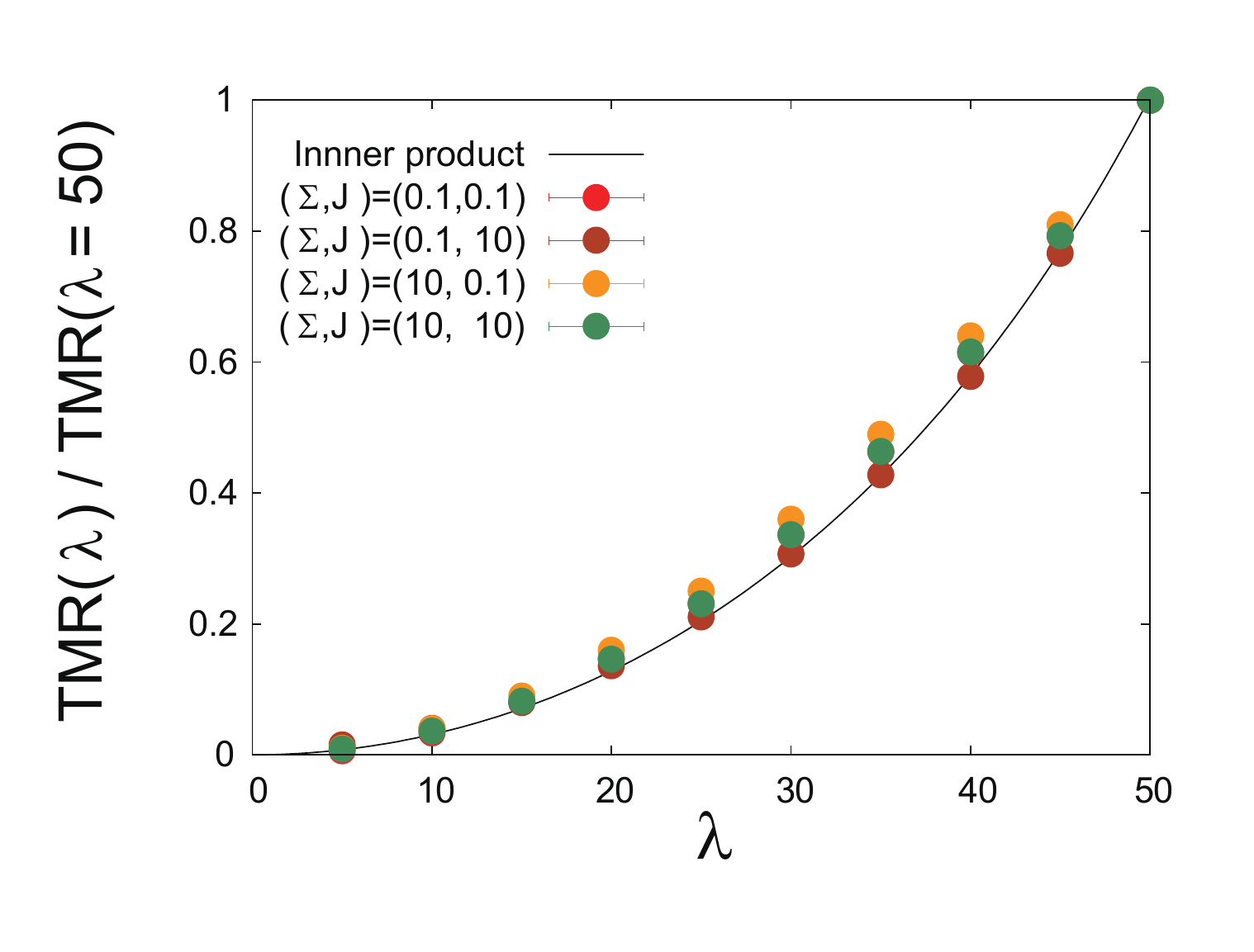}}
\caption{ (a) The skyrmion radius $\lambda$ dependence of the TMR ratio for various values of $\Sigma$ and $J$. All the points are well coincide with the estimation from the inner product of the local magnetic moments (black line).}
\label{FigRskdep}
\end{figure}

One expects that the TMR ratio would decrease from $30\%$ for smaller skyrmion. Fig.~\ref{FigRskdep} shows the skyrmion radius $\lambda$ dependence of the TMR ratio. If we adopt the simplest estimation from the inner product of the local magnetic moments, the TMR ratio can be estimated as $\mathrm{TMR}= \frac{\pi \lambda^2_{\mathrm{eff}}}{L^2-\pi \lambda^2_{\mathrm{eff}}}$ where $\lambda_{\mathrm{eff}}$, being defined as $\pi \lambda_{\mathrm{eff}}^2 \equiv \sum_{i}\frac{1-\vec{n}_i^{sk}\cdot\vec{n}_i^{F}}{2} =\sum_{i}\sin^2\frac{\theta_i}{2} = \pi \lambda^2 (\frac{1}{2}-\frac{2}{\pi^2})$, is the radius of a magnetic bubble whose effective number of flipped spins is the same as that of the present skyrmion profile. As seen in Fig.~\ref{FigRskdep}, the radius dependence of the TMR ratio is well reproduced by this simplest estimation for quite wide range of parameters even out of the plateau region in Fig.~\ref{FigSigmaJdep}(b). This result indicates that we have to fabricate the reading electrode of the MTJ devices as small as the diameter of the skyrmion to obtain the maximum value of TMR ratio.

The tunneling conductance in the skyrmionic systems are investigated in previous studies~\cite{crum2015,hanneken2015} where LDoS is modulated due to the hybridization of the spin majority and the minority states at the center of a skyrmion. These studies bear in mind the detection of a skyrmion using STM or STS, therefore the setup is different. In the STM, the tip can probe the surface of a device in the atomic resolution, and the tunneling conductance is proportional to LDoS at the tip position. However, in our set up, since the detection electrode (ferromagnetic layer) has the finite size, the in-plane propagation of the wave function in the electrode has a vital role. In our calculation, the tunnel current is written in terms of the Green function as $I\sim \mathrm{\ Tr}[\mathrm{Im } \left( \mathcal{\hat{G}}^{sk} \right)\mathcal{\hat{H}}_T\mathrm{Im} \left( \mathcal{\hat{G}}^F \right) \mathcal{\hat{H}}^\dagger_T ]$. If we take the trace in the real space, the tunneling Hamiltonian is nothing but $\left( H_T\right)_{ij}=T\delta_{ij}$, hence $I\sim|T|^2  \sum_{ij} [\mathrm{Im}\left( \mathcal{G}^{sk}_{ij} \right) \mathrm{Im} \left(  \mathcal{G}^F_{ji} \right)  ] $. If we neglect the in-plane propagation of electrons in the ferromagnetic layer and assume the homogeneity; $\mathcal{G}^F_{ji} = \mathcal{G}^F_0 \delta_{ij} $, the tunnel current can be written as the sum of LDoS of the skyrmion layer. This situation is true when the detection electrode is the array of isolated STM tips. But as one can see in Fig.~\ref{FigMudep}, estimations from DoS and LDoS are totally different from the real value. This result indicates the LDoS description is not enough when the detection electrode has the finite size as in the MTJ geometry.

%{\it Discussion}: 
We have demonstrated that a single skyrmion can be electrically detected using the TMR in the MTJ geometry.
%If only the spectral broadening $\Sigma$ is smaller than the exchange coupling $J$, the TMR ratio becomes larger than $30\%$. This value is sorely determined by the spin profile of the skyrmion regardless of the electronic structures. 
If only the spectral broadening $\Sigma$ is smaller than the exchange coupling $J$, the TMR ratio becomes larger than the value estimated sorely from the spin profile of the skyrmion. In the case of the maximum-size skyrmion in the square-shaped device, this value can amount to $30\%$.
Since this criterion $\Sigma \lesssim J$ is easily accessible in real systems, our proposed setup can be a good candidate for the reading part of the future skyrmionics memory devices.

This work was supported in
part by JSPS KAKENHI Grants No. 24224009, No. 25400317, No. 15H05854, and 
No. 26103006. 
K.H. is  supported by the Japan Society for the Promotion
of Science through a research fellowship for young scientists
and the Program for Leading Graduate Schools (MERIT).

\clearpage

\setcounter{figure}{0}
\setcounter{equation}{0}
\setcounter{table}{0}
\setcounter{page}{1}
\renewcommand{\thefigure}{S\arabic{figure}}
\renewcommand{\theequation}{S\arabic{equation}}
\renewcommand{\thetable}{S\arabic{table}} 
\renewcommand{\thesection}{S\arabic{section}} 

\begin{widetext}

%\twocolumn[

\section{Perturbation with respect to $J$}

In this section, we calculate the TMR ratio in the perturbative approach with respect to the exchange coupling $J$. We show $\mathrm{TMR} \propto J^2$ a shown in Fig.2(a) in the main text.

In the absence of $J$, the system is nearest neighbor tight-binding model on the $N\times N$ square lattice, therefore eigen vectors and energies are
\be
\psi_n^{(0)} (i) = \sqrt{\frac{2}{N+1}} \sin \frac{n\pi i}{N+1} , \qquad E_n^{(0)} = -2t\cos \frac{n\pi}{N+1} \qquad (n=1,...,N)
\ee
with the orthonormal condition;
\be
\sum_{i=1}^N \psi_m^{(0)\ast} (i) \psi_n^{(0)} (i) = \delta_{mn}.
\ee
As a perturbation, we consider the exchange coupling to the $z$ component of the local spin moment $n_z(i)$ for simplicity
\be
V(i) = -J n_z(i) \sigma_z .
\ee
For the skyrmion layer, $n^{sk}_z(i) = \cos\frac{2\pi i}{N+1}$ while that for the ferromagnetic layer is $n^F_z(i) = 1 $. Hereafter we consider only the spin-up electron, where the result for the spin down electron is obtained by the substitution $J\rightarrow -J$. In this case, the perturbation is just a local potential on the spin polarized electron. As the perturbation in the ferromagnetic layer is constant and diagonal, the effect of the perturbation is just a energy shift by $-J$.  The matrix element of the perturbation in the skyrmion layer is 
\begin{align}
V_{mn} &= \sum_i \psi_m^{(0)\ast} (i) V(i) \psi_n^{(0)} (i) =\frac{-2J}{N+1} \sum_i \sin \frac{m\pi i}{N+1} \sin \frac{n\pi i}{N+1} \cos\frac{2\pi i}{N+1} \\
&=-\frac{J}{2} (\delta _{m,n+2} + \delta _{m,n-2}).
\end{align}
Note that the system is homogeneous in $y$ direction therefore the wave function for the coordinate $y$ is the unperturbed one. The Dyson equation for the Matsubara Green function up to the order of $J^2$ reads
\be
G=G_0 + G_0VG_0 + G_0VG_0VG_0 + o(J^3)
\ee
where $(G_0)_{mn} =\delta_{mn}G_{0m}=\delta_{mn}\left[  i\omega - \xi_n +i \Gamma\mathrm{sgn}{\omega} \right]^{-1}$ is the non-perturbative Green function with energy $\xi_n = E_n^{(0)} -\mu$ and the Matsubara frequency $\omega$.
For the first order in $J$,
\be
( G_0VG_0 ) _{mn} = G_{0m} V_{mn} G_{0n} = -\frac{J}{2} G_{0m}G_{0n}(\delta _{m,n+2} + \delta _{m,n-2}).
\ee
while for the second order,
\begin{align}
( G_0VG_0VG_0 ) _{mn} &= \sum_k G_{0m} V_{mk} G_{0k} V_{kn} G_{0n} \\
&= \frac{J^2}{4} \sum_k G_{0m} G_{0k} G_{0n} (\delta _{m,k+2} + \delta _{m,k-2})(\delta _{k,n+2} + \delta _{k,n-2}) \\
&=G_{0m} G_{0m-2} G_{0n} (\delta _{m-2,n+2} + \delta _{m,n})  +  G_{0m} G_{0m+2} G_{0n} (\delta _{m,n} + \delta _{m+2,n-2})
\end{align}

The tunnel current is written as 
\begin{align}
I &= -2e \left. \mathrm{Im} U(i\omega) \right|_{i\omega \rightarrow eV+i\delta } \\
U(i\omega) &= \frac{1}{\beta}\sum_{i\Omega}  \mathrm{Tr} \left[ H_T G^1(i\Omega -i\omega) H_T G^2 (i\Omega )\right]
\end{align}
where $G^i$ is the perturbed Green function for $i$-th layer and $\Omega$ is the fermionic Matsubara frequency. By taking the trace using the unperturbed wave functions, the tunnel Hamiltonian and the ferromagnetic Hamiltonian is diagonal therefore, we obtain
\begin{align}
U(i\omega) &= |T|^2 \frac{1}{\beta}\sum_{i\Omega,n}    G^F_n (i\Omega -i\omega) G^{sk}_n (i\Omega )  \\
&=|T|^2\frac{1}{\beta}\sum_{i\Omega,n}    G^F_n (i\Omega -i\omega) \left[ G_{0n} (i\Omega )  + \frac{J^2}{4} (G_{0n}^2G_{0n-2}+G_{0n}^2G_{0n+2}) \right]\\
&=|T|^2\frac{1}{\beta}\sum_{i\Omega,k}    G^F_k (i\Omega -i\omega) \left[ G_{0k} (i\Omega )  + \frac{J^2}{2} G_{0k}^3 (i\Omega )  \right]\\
&=|T|^2\frac{1}{\beta}\sum_{i\Omega,k}    G^F_k (i\Omega -i\omega) \left[ 1+ \frac{J^2}{4} \frac{\partial^2}{\partial \mu^2}\right]G_{0k} (i\Omega ) .
\end{align}
Here we assumed the large system where we can neglect the difference between $n$ and $n\pm 2$ and moved on to the momentum representation. By the aid of the spectral representation;
\be
A(x) = -2 \mathrm{Im} G(x) , \qquad G(x) = \int \frac{A(x')}{x-x'+i\delta} \frac{\td x'}{2\pi}
\ee
the Matsubara summation for the zeroth order term is evaluated as
\begin{align}
U^{(0)}(i\omega)&= |T|^2\frac{1}{\beta}\sum_{i\Omega,k}    G^F_k (i\Omega -i\omega) G_{0k} (i\Omega )  \\
&=   |T|^2\frac{1}{\beta}\sum_{i\Omega,k} \int \frac{\td x}{2\pi} \frac{\td y}{2\pi} \frac{A^F_k(x)}{i\Omega -i\omega-x+i\delta} \frac{A_{0k}(x)}{i\Omega -y+i\delta} \\
&=  |T|^2\sum_k\int \frac{\td x}{2\pi} \frac{\td y}{2\pi}  \frac{f(x)-f(y)}{i\omega+x-y}A^F_k(x)A_{0k}(y).
\end{align}
After the analytical continuation, 
\begin{align}
I^{(0)}&=-2e\mathrm{Im} U^{(0)}(i\omega) \left. \right|_{i\omega \rightarrow eV+i\delta } \\
&= -2e |T|^2 \mathrm{Im} \sum_k\int \frac{\td x}{2\pi} \frac{\td y}{2\pi}  \frac{f(x)-f(y)}{eV+x-y +i\delta}A^F_k(x)A_{0k}(y).  \\
&= 2e |T|^2 \sum_k\int \frac{\td x}{2\pi} \frac{\td y}{2\pi}  \left[f(x)-f(y)\right] \pi\delta(eV+x-y)A^F_k(x)A_{0k}(y).  \\
&= e |T|^2 \sum_k\int  \frac{\td x}{2\pi}  \left[f(x)-f(x+eV)\right] A^F_k(x)A_{0k}(x+eV).  \\
&= \frac{e^2V }{2\pi}|T|^2 \sum_k    A^F_k(0)A_{0k}(0).  \\
&= \frac{e^2V }{2\pi}|T|^2 \sum_k  \frac{2\Sigma}{(E^{(0)}_k-J-\mu)^2+\Sigma^2}\frac{2\Sigma}{(E^{(0)}_k-\mu)^2+\Sigma^2}  \\
&= \frac{e^2V }{2\pi}|T|^2 D_0 \int_{-\infty}^{\infty}\td E \frac{2\Sigma}{(E-J-\mu)^2+\Sigma^2}\frac{2\Sigma}{(E-\mu)^2+\Sigma^2}  \\
&= \frac{e^2V }{2\pi}|T|^2 D_0  \frac{8\pi \Sigma}{4\Sigma^2 + J^2}
\end{align}
where we assumed the zero temperature and the linear response. $D_0$ is the density of states at the Fermi energy which is assumed to be constant. The second order one is calculated as
\begin{align}
I^{(2)}&= \frac{e^2V }{2\pi}|T|^2 D_0 \int_{-\infty}^{\infty}\td E \frac{2\Sigma}{(E-J-\mu)^2+\Sigma^2} \frac{J^2}{4} \frac{\partial^2}{\partial \mu^2}  \frac{2\Sigma}{(E-\mu)^2+\Sigma^2}  \\
&= \frac{e^2V }{2\pi}|T|^2 D_0 \frac{-\pi J^2}{4\Sigma^3}.
\end{align}

The tunnel current in the parallel ferromagnetic configuration is easily calculated as
\begin{align}
I_P&= \frac{e^2V }{2\pi}|T|^2 D_0 \int_{-\infty}^{\infty}\td E \frac{2\Sigma}{(E-J-\mu)^2+\Sigma^2} \frac{2\Sigma}{(E-J-\mu)^2+\Sigma^2}  \\
&= \frac{e^2V }{2\pi}|T|^2 D_0 \frac{2\pi}{\Sigma}.
\end{align}
The current for the spin down electron, which is obtained by the substitution $J\rightarrow -J$, is the same as that for the spin up electron since all the currents are the even function of $J$ Thus, the TMR ratio is
\be
\mathrm{TMR} = \frac{I_P}{I^{(0)}+I^{(2)}} -1 = \frac{3J^2}{8\Sigma^2} + o(J^4)
\ee.

\section{TMR for antiparallel configuration}
In this section, we analytically derive the TMR ratio in the antiparallel ferromagnetic bilayer system and show $\mathrm{TMR} \propto \Sigma^{-2}$ in the clean limit. This is the indirect evidence that $\mathrm{TMR} \propto \Sigma^{-2}$ holds even in the skyrmion-ferromagnet bilayer system.

The tunneling current in the bilayer system is written as
\be
\displaystyle I = \frac{e^2V}{2\pi}\sum_{mn}\left|T_{mn}\right|^2 \frac{2\Sigma}{(\varepsilon^1_m-\mu)^2+\Sigma^2}\frac{2\Sigma}{(\varepsilon^2_n-\mu)^2+\Sigma^2}
\ee
where $\mu$ is the chemical potential, $\Sigma$ is the spectral broadening, $\varepsilon^i_m$ is the $m$-th eigen energy of the layer $i$, and $T_{mn}$ is the tunneling matrix element in the eigen basis which accounts for the spin-conserving vertical hopping in the real space. Hereafter, we assume the system is the homogeneous ferromagnet with translational invariance. In this condition, the tunneling matrix is $\left|T_{mn}\right|^2\rightarrow \left|T\right|^2 \delta_{kk'} \delta_{\sigma\sigma'} $. The expression for the tunneling current is
\be
\displaystyle I = \frac{e^2V}{2\pi}\left|T\right|^2\sum_{k\sigma} \frac{2\Sigma}{(\varepsilon^1_{\bm{k}\sigma}-\mu)^2+\Sigma^2}\frac{2\Sigma}{(\varepsilon^2_{\bm{k}\sigma}-\mu)^2+\Sigma^2}
\ee
which shows the spin, the momentum, and the energy is preserved during the tunneling process if $\Sigma$ goes zero. In the parallel configuration, the current $I_P$ is calculated using $\varepsilon^1_{\bm{k}\sigma}=\varepsilon^2_{\bm{k}\sigma}=\varepsilon^0_{\bm{k}}-J\sigma$ while we use $\varepsilon^1_{\bm{k}\sigma}=\varepsilon^0_{\bm{k}}-J\sigma$ and $\varepsilon^2_{\bm{k}\sigma}=\varepsilon^0_{\bm{k}}+J\sigma$ for the antiparallel current $I_{AP}$. For $\varepsilon^0_{\bm{k}}$, the energy without exchange coupling, we assume the parabollic dispersion with constant density of states $D_0$. This assumption is justified if the distances between the chemical potential and the band bottom or the van Hove singularities are much larger than the spectral broadening $\Sigma$. In this situation, the integral over the energy can be extended from the band width to the infinity since the integrand decays suddenly for small $\Sigma$. The current in each configuration is written as
\begin{align}
I_P &= \frac{e^2V}{2\pi}\left|T\right|^2 D_0 \int_{-\infty}^{\infty} \td \varepsilon \sum_{\sigma} \frac{2\Sigma}{(\varepsilon-J \sigma -\mu)^2+\Sigma^2}\frac{2\Sigma}{(\varepsilon-J\sigma-\mu)^2+\Sigma^2} \\
&= \frac{e^2V}{2\pi}\left|T\right|^2 D_0 \frac{2\pi}{\Sigma} \times 2 \\
I_{AP} &= \frac{e^2V}{2\pi}\left|T\right|^2 D_0 \int_{-\infty}^{\infty} \td \varepsilon \sum_{\sigma} \frac{2\Sigma}{(\varepsilon-J \sigma -\mu)^2+\Sigma^2}\frac{2\Sigma}{(\varepsilon+J\sigma-\mu)^2+\Sigma^2} \\
&= \frac{e^2V}{2\pi}\left|T\right|^2 D_0 \frac{2\pi\Sigma}{\Sigma^2+J^2} \times 2.
\end{align}
Note that the tunneling current in the antiparallel configuration vanishes in the clean limit since the spin, the momentum, and the energy are conserved values during the tunneling process. The TMR ratio is
\be
TMR=\frac{I_P}{I_{AP}}-1 = \frac{\Sigma^2 + J^2}{\Sigma^2} -1 = \frac{J^2}{\Sigma^2}
\ee
which shows the TMR ratio diverges as $TMR\propto \Sigma^{-2}$ as $\Sigma \rightarrow 0$.

For the general value of $\Sigma$, The integral is from $-4t$ to $4t$, therefore we obtain
\begin{align}
I_P&= \frac{e^2V}{2\pi}\left|T\right|^2 D_0\frac{2}{\Sigma } \left[\Sigma \left(
\frac{4t-J-\mu}{(4t-J-\mu)^2+\Sigma ^2}
+\frac{4t+J-\mu}{(4t+J-\mu)^2+\Sigma ^2}
+\frac{4t-J+\mu}{(4t-J+\mu)^2+\Sigma^2}
+\frac{4t+J+\mu}{(4t+J+\mu)^2+\Sigma ^2}\right) \right.  \nonumber \\
& \left. 
+\tan ^{-1}\left(\frac{4t-J-\mu}{\Sigma }\right)
+\tan ^{-1}\left(\frac{4t+J-\mu}{\Sigma }\right)
+\tan ^{-1}\left(\frac{4t-J+\mu}{\Sigma }\right)
+\tan ^{-1}\left(\frac{4t+J+\mu}{\Sigma }\right) \right] \\
I_{AP} &= \frac{e^2V}{2\pi}\left|T\right|^2 D_0 \frac{\Sigma}{J \left(J^2+\Sigma^2\right)} 
\left[ \Sigma  \left(
 \log \left( \frac{ (4t + J-\mu)^2+\Sigma ^2} { (4t -J-\mu)^2+\Sigma ^2} \right)
+\log \left( \frac{ (4t + J+\mu)^2+\Sigma ^2} { (4t-J+\mu )^2+\Sigma ^2}\right)\right)  \right.  \nonumber \\
& \left.  +2 J \left(
+\tan ^{-1}\left(\frac{4t-J+\mu}{\Sigma }\right)
+\tan ^{-1}\left(\frac{4t+J-\mu}{\Sigma }\right)
+\tan ^{-1}\left(\frac{4t-J+\mu}{\Sigma }\right)
+\tan ^{-1}\left(\frac{4t+J+\mu}{\Sigma }\right) \right) \right]
\end{align}
TMR behaves as
\be
TMR = \begin{cases}
    \frac{8 J^2}{3 \Sigma ^4} \left(3 \mu ^2+16 t^2\right)+O\left(\Sigma^{-5}\right) &   (\rm{Large}\  \Sigma) \\
    \frac{J^2}{\Sigma^2} +O\left(\Sigma^{-1}\right) & (\rm{Small}\  \Sigma)
  \end{cases} 
\ee
Although this is the result for the antiparallel ferromagnet-ferromagnet system, it well captures limiting behaviors of the skyrmion-ferromagnetic bilayer system shown in the Fig.2(b) in the main text. We also have numerically confirmed that these behaviors do not change even if we take into account the realistic density of states of the square lattice, which contain the van Hove singularity at $\varepsilon=0$.

\section{Large $\Sigma$ region}
In this section, we show $\mathrm{TMR}\propto\Sigma^{-4}$ in the skyrmion-ferromagnet bilayer system for large $\Sigma$ as depicted in the Fig.2(b) in the main text. From the eq.(6) in the main text, the tunneling current for large $\Sigma$ is expanded as
\begin{align}
I_{sk} &= \frac{2e^2V}{\pi}\Sigma^{-2} \sum_{mn}\left|T_{mn}\right|^2 \left[ 1 + \left(\frac{\varepsilon^F_m-\mu}{\Sigma}\right)^2 \right]^{-1}\left[ 1 + \left( \frac{\varepsilon^{sk}_n-\mu}{\Sigma}\right)^2 \right]^{-1} \nonumber\\
 &= \frac{2e^2V}{\pi}\Sigma^{-2} \sum_{mn}\left|T_{mn}\right|^2 \left[ 1 - \left(\frac{\varepsilon^F_m-\mu}{\Sigma}\right)^2 - \left( \frac{\varepsilon^{sk}_n-\mu}{\Sigma}\right)^2 \right] +o(\Sigma^{-6}) \nonumber \\
 &= \frac{4e^2V}{\pi}\Sigma^{-2} L^2 \left[ 1 - \frac{1}{2L^2}\sum_n \left[ \left(\frac{\varepsilon^F_n-\mu}{\Sigma}\right)^2 + \left( \frac{\varepsilon^{sk}_n-\mu}{\Sigma}\right)^2 \right] \right] +o(\Sigma^{-6}).
\end{align}
Here we have used $\sum_n \left|T_{mn}\right|^2 =\sum_m \left|T_{mn}\right|^2 =1$.
The TMR ratio is
\begin{align}
\mathrm{TMR} &= \frac{I_P}{I_{sk}}-1 \nonumber\\
&= \frac{ 1 - \frac{1}{2L^2}\sum_n \left[ \left(\frac{\varepsilon^F_n-\mu}{\Sigma}\right)^2 + \left( \frac{\varepsilon^{F}_n-\mu}{\Sigma}\right)^2 \right] +o(\Sigma^{-4})}{ 1 - \frac{1}{2L^2}\sum_n \left[ \left(\frac{\varepsilon^F_n-\mu}{\Sigma}\right)^2 + \left( \frac{\varepsilon^{sk}_n-\mu}{\Sigma}\right)^2 \right] +o(\Sigma^{-4})} -1 \nonumber\\
&= \frac{1}{2L^2}\sum_n \left[ \left(\frac{\varepsilon^{sk}_n-\mu}{\Sigma}\right)^2 - \left( \frac{\varepsilon^F_n-\mu}{\Sigma}\right)^2 \right]+o(\Sigma^{-4})  \nonumber\\
&= \frac{1}{2L^2\Sigma^2} \mathrm{Tr} \left[ (\hat{\mathcal{H}}_{sk} -\mu)^2 -(\hat{\mathcal{H}}_{F}-\mu)^2   \right]+o(\Sigma^{-4}) \nonumber \\
&= \frac{1}{2L^2\Sigma^2} \mathrm{Tr} \left[ (\hat{\mathcal{H}}_{sk} )^2 -(\hat{\mathcal{H}}_{F})^2   \right]+o(\Sigma^{-4}) .
\end{align}
since $\hat{\mathcal{H}}_{sk}$ and $\hat{\mathcal{H}}_{F}$ themselves are trace-less. That can be seen the hopping term is trace-less in the real space and the double exchange term is also in the spin space. We next separate the Hamiltonians into the hopping terms and the double exchange terms as $\hat{\mathcal{H}}_{F} = 
\hat{\mathcal{K}} + \hat{\mathcal{J}}_F$ and  $\hat{\mathcal{H}}_{sk} = \hat{\mathcal{K}} + \hat{\mathcal{J}}_{sk}$
\begin{align}
&  \mathrm{Tr} \left[ (\hat{\mathcal{H}}_{sk} )^2 -(\hat{\mathcal{H}}_{F})^2   \right]  \nonumber \\
&=\mathrm{Tr} \left[ 2\hat{\mathcal{K}} (\hat{\mathcal{J}}_{sk}-\hat{\mathcal{J}}_{sk} ) +\hat{\mathcal{J}}_{sk}^2-\hat{\mathcal{J}}_{F}^2  \right] .
\end{align}
The first term vanishes since $\hat{\mathcal{K}}$ and $\hat{\mathcal{J}}$ act on different space; $\mathrm{Tr} \left[ \hat{\mathcal{K}} \otimes \hat{\mathcal{J}} \right] = \mathrm{Tr} \hat{\mathcal{K}} \times \mathrm{Tr}  \hat{\mathcal{J}} =0 $. The second and the term cancel each other as $ \hat{\mathcal{J}}^2 = \left(-J\vec{\sigma} \cdot\vec{n}\right)^2 = J^2 $ is independent of the spin configurations $\vec{n}$. In summary, order of $\Sigma^{-2}$ vanishes hence $\mathrm{TMR}\propto\Sigma^{-4}$.

\end{widetext}
%]


\begin{thebibliography}{99}
\bibitem{nagaosa2013}N. Nagaosa, and Y. Tokura,
% “Topological properties and dynamics of magnetic skyrmions.,?�?�?
Nat. Nanotechnol., \textbf{8}, 899, (2013).

\bibitem{fert2017}A. Fert, N. Reyren, and V. Cros
%Magnetic skyrmions: advances in physics and potential applications
Nat. Rev. Mat., \textbf{2}, 17031, (2017).

\bibitem{muhlbauer2009}S. M\"{u}hlbauer, B. Binz, F. Jonietz, C. Pfleiderer, A. Rosch, A. Neubauer, R. Georgii, and P. B\"{o}ni,
%“Skyrmion Lattice in a Chiral Magnet,?�?�?
Science, \textbf{323}, 915, (2009).

\bibitem{yu2010}X. Z. Yu, Y. Onose, N. Kanazawa, J. H. Park, J. H. Han, Y. Matsui, N. Nagaosa, and Y. Tokura, 
%“Real-space observation of a two-dimensional skyrmion crystal.,?�?�?
Nature, \textbf{465}, 901 (2010).

\bibitem{okubo2012}T. Okubo, S. Chung, and H. Kawamura, 
%“Multiple- q States and the Skyrmion Lattice of the Triangular-Lattice Heisenberg Antiferromagnet under Magnetic Fields,?�?�?
Phys. Rev. Lett., \textbf{108}, 017206, (2012).

\bibitem{hayami2016}S. Hayami, S. Lin, and C. D. Batista, 
%“Bubble and skyrmion crystals in frustrated magnets with easy-axis anisotropy,?�?�?
Phys. Rev. B, \textbf{93}, 184413, (2016).

\bibitem{lin2016}S. Lin and S. Hayami, 
%“Ginzburg-Landau theory for skyrmions in inversion-symmetric magnets with competing interactions,?�?�?
Phys. Rev. B, \textbf{93}, 064430, (2016).


\bibitem{matsuno2016}J. Matsuno, N. Ogawa, K. Yasuda, F. Kagawa, W. Koshibae, and N. Nagaosa, 
%“Interface-driven topological Hall effect in,?�?�?
Sci. Adv., \textbf{2}, e1600304 (2016).

\bibitem{dupe2016} B. D\'{u}pe, G. Bihlmayer, M. Bottcher, S. Bl\"{u}gel, and S. Heinze, 
%“Engineering multilayers for spintronics,?�?�?
Nat. Commun., \textbf{7}, 11779, (2016).

\bibitem{jonietz2010}F. Jonietz, S. M\"{u}hlbauer, C. Pfleiderer, A. Neubauer, W. M\"{u}nzer, A. Bauer, T. Adams, R. Georgii, P. B\"{o}ni, R. a Duine, K. Everschor, M. Garst, and a Rosch, 
%“Spin transfer torques in MnSi at ultralow current densities.,?�?�?
Science, \textbf{330}, 1648, (2010).

\bibitem{iwasaki2013a}J. Iwasaki, M. Mochizuki, and N. Nagaosa, 
%“Universal current-velocity relation of skyrmion motion in chiral magnets.,?�?�?
Nat. Commun., \textbf{4}, 1463, (2013).

\bibitem{heinze2011}S. Heinze, K. Von Bergmann, M. Menzel, J. Brede, A. Kubetzka, R. Wiesendanger, G. Bihlmayer, and S. Bl\"{u}gel, 
%“Spontaneous atomic-scale magnetic skyrmion lattice in two dimensions,?�?�?
Nat. Phys., \textbf{7}, 713, (2011).

\bibitem{koshibae2015}W. Koshibae, Y. Kaneko, J. Iwasaki, M. Kawasaki, Y. Tokura, and N. Nagaosa, 
%“Memory functions of magnetic skyrmions,?�?�?
 Jpn. J. Appl. Phys., \textbf{54}, 053001, (2015).

\bibitem{parkin2008} S. S. P. Parkin, M. Hayashi, and L. Thomas,
%Magnetic Domain-Wall Racetrack Memory
Science, \textbf{320}, (5873), 190, (2008).

\bibitem{kang2016}B. W. Kang, Y. Huang, X. Zhang, Y. Zhou, and W. Zhao, 
%“Skyrmion-Electronics : An Overview and Outlook,?�?�?
IEEE Proc., \textbf{104}, 2040, (2016).

\bibitem{sampaio2013}J. Sampaio, V. Cros, S. Rohart, A. Thiaville, and A. Fert, 
%“Nucleation, stability and current-induced motion of isolated magnetic skyrmions in nanostructures,?�?�?
Nat. Nanotechnol., \textbf{8}, 839 (2013).

\bibitem{romming2013}N. Romming, C. Hanneken, M. Menzel, J. E. Bickel, B. Wolter, K. von Bergmann, A. Kubetzka, and R. Wiesendanger, 
%“Writing and Deleting Single Magnetic Skyrmions,?�?�?
Science, \textbf{341}, 636, (2013).

\bibitem{finazzi2013}M. Finazzi, M. Savoini, A. R. Khorsand, A. Tsukamoto, A. Itoh, L. Du\`{o}, A. Kirilyuk, and M. Ezawa, 
%“Laser-Induced Magnetic Nanostructures with Tunable Topological Properties,?�?�?
Phys. Rev. Lett., \textbf{110}, 177205, (2013).

\bibitem{koshibae2014} W. Koshibae, and N. Nagaosa
%Creation of skyrmions and antiskyrmions by local heating
Nat. Commun., \textbf{5}, 5148, (2014).

\bibitem{iwasaki2013b}J. Iwasaki, M. Mochizuki, and N. Nagaosa,
% “Current-induced skyrmion dynamics in constricted geometries,?�?�?
Nat. Nanotechnol., \textbf{8}, 742, (2013).

\bibitem{hanneken2015}C. Hanneken, F. Otte, A. Kubetzka, B. Dup\'{e}, N. Romming, K. von Bergmann, R. Wiesendanger, and S. Heinze, 
%“Electrical detection of magnetic skyrmions by tunnelling non-collinear magnetoresistance,?�?�?
Nat. Nanotechnol., \textbf{10}, 1039, (2015).

\bibitem{crum2015}D. M. Crum, M. Bouhassoune, J. Bouaziz, and B. Schweflinghaus, S. Bl\"{u}gel, and S. Lounis
%“Perpendicular reading of single confine magnetic skyrmions,?�?�?
Nat. Commun., \textbf{6}, 8541, (2015).

\bibitem{kanazawa2015}N. Kanazawa, M. Kubota, a. Tsukazaki, Y. Kozuka, K. S. Takahashi, M. Kawasaki, M. Ichikawa, F. Kagawa, and Y. Tokura, 
%“Discretized topological Hall effect emerging from skyrmions in constricted geometry,?�?�?
 Phys. Rev. B, \textbf{91}, 041122(R), (2015).

\bibitem{hamamoto2016}K. Hamamoto, M. Ezawa, N. Nagaosa, K. Hamamoto, M. Ezawa, and N. Nagaosa, 
%“Purely electrical detection of a skyrmion in constricted geometry,?�?�?
Appl. Phys. Lett., \textbf{108}, 112401, (2016).


%%%%%%%%intro%%%%%%%

\bibitem{mahan} G. D. Mahan, \textit{Many-Particle Physics}, 3rd. ed. (Kluwer Academic/Plenum Publishers, New York, 2000).

\bibitem{maekawa1982}S. Maekawa and U. Gafvert, 
%“Electron Tunneling Between Ferromagnetic Films,?�?�?
IEEE Trans. Electron Devices, \textbf{MAG-18}, 707, (1982).


\end{thebibliography}
\end{document}